# Domestic Activities Classification from Audio Recordings Using Multi-scale Dilated Depthwise Separable Convolutional Network

Yufei Zeng
School of Electronic and Information Engineering, South China University of Technology, Guangzhou, China
1013990615@qq.com

Yanxiong Li*
School of Electronic and Information Engineering, South China University of Technology, Guangzhou, China
eeyxli@scut.edu.cn

Zhenfeng Zhou
School of Electronic and Information Engineering, South China University of Technology, Guangzhou, China
310378072@qq.com

Ruiqi Wang
School of Electronic and Information Engineering, South China University of Technology, Guangzhou, China
1063749117@qq.com

Difeng Lu
School of Electronic and Information Engineering, South China University of Technology, Guangzhou, China
792280665@qq.com

*Abstract*—Domestic activities classification (DAC) from audio recordings aims at classifying audio recordings into pre-defined categories of domestic activities, which is an effective way for estimation of daily activities performed in home environment. In this paper, we propose a method for DAC from audio recordings using a multi-scale dilated depthwise separable convolutional network (DSCN). The DSCN is a lightweight neural network with small size of parameters and thus suitable to be deployed in portable terminals with limited computing resources. To expand the receptive field with the same size of DSCN's parameters, dilated convolution, instead of normal convolution, is used in the DSCN for further improving the DSCN's performance. In addition, the embeddings of various scales learned by the dilated DSCN are concatenated as a multi-scale embedding for representing property differences among various classes of domestic activities. Evaluated on a public dataset of the Task 5 of the 2018 challenge on Detection and Classification of Acoustic Scenes and Events (DCASE-2018), the results show that: both dilated convolution and multi-scale embedding contribute to the performance improvement of the proposed method; and the proposed method outperforms the methods based on state-of-the-art lightweight network in terms of classification accuracy.

*Keywords—Domestic activities classification, multi-scale embedding, dilated convolution, depthwise separable convolution*

## I. Introduction

With the development of economy and society, researchers have been making great efforts on enhancing the living quality for humans in terms of safety, comfort, and home care [1]. The classification of domestic activities from audio recordings has become a new trend in the smart-home field. Considering the characteristics of home environment (such as many light-blocking objects and residents' privacy), the video-based surveillance technique is not a reasonable solution for domestic activities classification. On the contrary, the audio-based surveillance technique has the advantages of low computational load, small storage space, no monitoring blind area and privacy protection to a certain extent.

In this study, we focus on classifying domestic activities from audio recordings for audio surveillance in home environment. This work is motivated by a practical audio surveillance application for solving the nursing problems of the elderly living alone in China where the problem of population aging is very serious. The goal of our work is to realize a DAC system in portable terminal with very limited computing resources which is deployed in home environment.

From a technical perspective, the task of DAC can be seen as a problem of acoustic scene classification (ASC). An acoustic scene generally consists of some relevant sound events which occur together. The problem of ASC is still challenging and far from being tackled, because the properties of various classes of acoustic scenes are diverse and complex with different time-frequency resolutions [2], [3]. As a result, single-scale embedding (SCE) learned by a deep neural network might be not enough to achieve better performance for ASC. To effectively represent the property differences among various classes of acoustic scenes, multi-scale embedding (MSE) learned by a dilated DSCN is adopted in this work.

In the proposed method, log Mel-spectrogram is used as the input audio feature of the dilated DSCN for learning MSE due to its excellent performance in audio processing tasks. The DSCN is a lightweight network which is able to effectively overcome the overfitting problem [4], and is thus suitable to be deployed in portable terminals with limited computing resources. That is, the DSCN is able to meet the requirement of our work for portable terminal application. In addition, the dilated convolutional kernel [5] can expand the receptive field without increasing the size of network's parameters. Therefore, to gain the same size of receptive field, the DSCN with dilated convolution needs fewer parameters than the DSCN with normal convolution, and is able to reduce the spatial complexity of the neural network. It has been proved that long temporal context information modelled by the dilated convolutions is effective for improving the performance of neural networks in some related tasks. For example, Li et al [6] proposed to use a dilated convolutional recurrent neural network (CRNN) for modeling the long temporal context of sound events, and their dilated CRNN outperformed the CRNN without dilated convolutions.

* Corresponding author: Yanxiong Li (eeyxli@scut.edu.cn).

This work was partly supported by national natural science foundation of China (61771200, 62111530145), international scientific research collaboration project of Guangdong Province, China (2021A0505030003), Guangdong basic and applied basic research foundation, China (2021A1515011454), and the national undergraduate training program for innovation and entrepreneurship (202010561023).

In this paper, different DAC methods are evaluated using one public dataset of the Task 5 of DCASE-2018. In addition, we investigate the contributions of both MSE module and dilated convolutional operation to the performance improvement of the proposed method. Experimental results indicate that our method exceeds the methods based on state-of-the-art lightweight networks in terms of classification accuracy.

In summary, main contributions of this study are as follows. First, we propose a DAC method using MSE learned by a lightweight network, i.e., dilated DSCN. The proposed dilated DSCN inherits merits from some state-of-the-art models, such as aggregation of multiple SCEs, dilated convolution, and depthwise separable convolution. To the best of our knowledge, this study is the first work for investigating the application of lightweight network with MSE to tackle the problem of DAC. Second, we evaluate the capabilities of the proposed DAC method and state-of-the-art methods on one public dataset.

The rest of this paper is structured as follows. Related works are introduced in Section II, and the proposed method is described in Section III. Experiments and conclusions are presented in Sections IV and V, respectively.

## II. RELATED WORKS

ASC is a quite active research field in recent years. Many evaluation campaigns, such as the DCASE [7]-[9], have promoted the development of ASC techniques. According to the types of the features adopted in different works, the ASC methods are summarized as follows.

Many hand-crafted features and transformed features have been proposed for ASC. Hand-crafted features are first proved to be effective for ASC in many previous works. The popular hand-crafted features include: Log Mel-spectrogram, Gabor filterbank, linear prediction cepstral coefficients, spectral centroid magnitude cepstral coefficients, Mel frequency cepstral coefficients, Mel-frequency discrete wavelet coefficients, spectral flux, higher-order ambisonic features, histogram of gradients for time-frequency representations, local binary patterns, and hash features [9]-[20]. Although these hand-crafted features are effective in specific conditions, they are hard to be generalized to other experimental conditions in practical applications and thus lack a flexibility in practice.

In order to overcome the shortcomings of hand-crafted features above, some researchers design transformed features based on non-negative matrix factorization (NMF) [21], [22] and deep neural network (DNN) [23]-[29]. The DNN-based transformed features have obtained more satisfactory results for ASC than both hand-crafted and NMF-based features when sufficient training data are available [8]. In addition, the transformed embedding generated by one layer of a DNN is popularly used as the final transformed feature [24]-[29], and is called SCE here. Although the DNN-based features outperform other features, it is still insufficient to use SCE for effectively discriminating various acoustic scenes due to complex time-frequency properties of different acoustic scenes [30]. To further improve the ASC performance, MSE is learned by non-lightweight networks in some recent works [30]-[33]. However, lightweight neural networks are not considered for MSE learning in these previous works. Hence, the effectiveness of the lightweight neural networks for MSE learning needs to be investigated.

## III. METHOD

The framework of the proposed method is shown in Fig. 1, which mainly consists of four modules: input module, dilated depthwise separable convolution (DSC) module, MSE module, and output module.

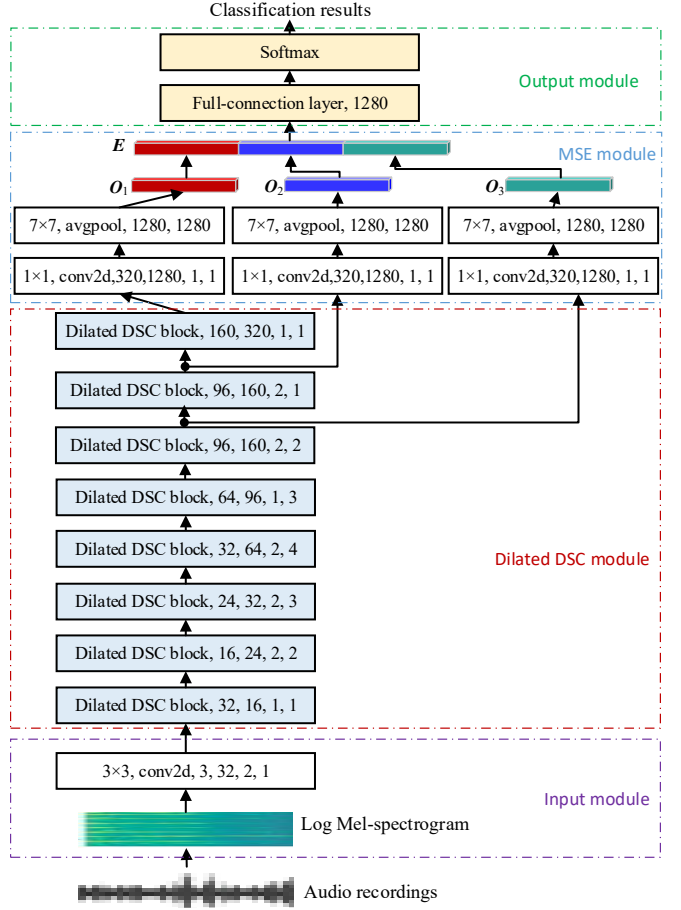

Fig. 1. Framework of the proposed method.

### A. Input Module

The input module includes the extraction of log Mel-spectrogram from audio recordings and a convolutional layer. Log Mel-spectrogram is an effective and popular acoustic features for ASC. Log Mel-spectrogram is first extracted from each audio recording, and then convolved with a 3×3 convolution kernel which is contained in a convolutional layer. In the convolutional layer (i.e., "3×3, conv2d, 3, 32, 2, 1"), 3×3, conv2d, 3, 32, 2, and 1 represent the size of convolutional kernel (3×3), two-dimensional convolution, the channel number of input feature map (3), the channel number of output feature map (32), the stride (2), and the number of repetitions for this convolutional layer (1 time), respectively.

### B. Dilated DSC Module

The dilated DSC module includes 8 dilated DSC blocks whose hyperparameters are listed in Fig. 1. In each DSC block (e.g., "Dilated DSC block, 32, 16, 1, 1"), 32, 16, 1, and 1 represent the channel number of input feature map (32), the channel number of output feature map (16), the stride (1), and the number of repetitions for this block (1 time), respectively. Each dilated DSC block consists of three dilated convolutional layers, and the parameters settings of these three convolutional layers in each dilated DSC block are identical.

The depthwise separable convolution is realized by decomposing the normal convolution into depthwise convolution and pointwise 1×1 convolution [34]. Depthwise convolution applies a single filter to each input channel, and pointwise convolution uses 1×1 convolution to combine the outputs of different depthwise convolutions. This decomposition process can greatly reduce the amount of calculation and the size of the network [34]. Batch normalization (BN) and rectified linear unit (ReLU) are adopted in each DSC block. Depthwise convolution can be written in the following form (one filter per input channel):

$$\hat{G}_{k,l,m} = \sum_{i,j} \hat{K}_{i,j,m} \cdot \hat{F}_{k+i-1,l+j-1,m} \quad (1)$$

where $\hat{K}$ denotes the convolutional kernel of depthwise convolution, whose size is $D_K \cdot D_K \cdot M$; $M$ stands for the number of input feature map channel; $D_k$ represents the size of the convolutional kernel. The $m$-th filter in $\hat{K}$ is applied to the $m$-th feature vector of the input feature map $F$ for producing the $m$-th channel of the output feature $G$. The computational complexity of depthwise convolution is: $D_k \times D_k \times M \times D_F \times D_F$, where $D_F$ denotes the size of the feature map. If floating-point operations are considered, the number of operations for a set of two-dimensional convolutional kernels to complete depthwise convolution is: $D_k \times D_k \times N \times M \times D_F \times D_F$, where $N$ denotes the channel number of output feature maps. The number of floating-point operations for pointwise convolution operation is: $N \times M \times D_F \times D_F$. In summary, the total amount of calculation for a depthwise separable convolution is: $D_k \times D_k \times M \times D_F + N \times M \times D_F \times D_F$.

To expand the receptive field and extract the contextual relationship of the acoustic scene in each audio recording, we adopt a dilated convolution for further improving the performance. A two-dimensional convolution operation is implemented by convolving the feature map $F$ with a convolutional kernel $K$ of size $(2n + 1) \times (2n + 1)$, whose definition is as follows:

$$(F * K)(q) = \sum_{q=s+t} F(s)K(t), \quad (2)$$

where $q, s \in \mathbb{Z}^2$, $t \in [-n, n]^2 \cap \mathbb{Z}^2$; $\mathbb{Z}$ represents the set of integers; $*$ stands for convolutional operator. The dilated convolutional operator $*_d$, is defined by

$$(F *_d K)(q) = \sum_{q=s+dt} F(s)K(t), \quad (3)$$

where $q, s \in \mathbb{Z}^2$, $t \in [-n, n]^2 \cap \mathbb{Z}^2$, $d$ represents the dilation rate, and $*_d$ represents a dilated convolution with dilation rate of $d$. Therefore, normal convolution can be considered as the dilated convolution with dilation rate of 1. In the same way, a one-dimensional dilated convolution with dilation rate of $d$ can be defined by Eq. (3), but its definition domain is different from the two-dimensional convolution. The definition domain of one-dimensional dilated convolution is: $q, s \in \mathbb{Z}$, $t \in [-n, n] \cap \mathbb{Z}$.

*C. Multi-scale Embedding Module*

As shown in Fig. 1, three embeddings (feature maps) outputted from the last three dilated DSC blocks are respectively fed to one convolutional layer (i.e., "1×1, conv2d, 320, 1280, 1, 1") and then respectively fed to one average-pooling layer (i.e., "7×7, avgpool, 1280, 1280") for further transformation. The meaning of each part of the "1×1, conv2d, 320, 1280, 1, 1" is the same to that of the counterpart in the convolutional layer in input module. In the average-pooling layer, 7×7, avgpool, 1280, and 1280 denote the size of pooling unit (7×7), average-pooling operation (avgpool), the channel number of input feature map (1280), and the channel number of output feature map (1280), respectively. The outputs of three average-pooling layers in Fig. 1, namely $O_1$, $O_2$ and $O_3$, are concatenated to obtain the MSE $E$. By the transformation of different layers, more semantic information can be further learned and the final embedding after concatenation is fed to the Softmax layer.

*D. Output Module*

The output module includes a full-connection layer and a Softmax layer, in which the three-scale embedding $E$ is first fed to the full-connection layer for further transformation, and then used as the input of the Softmax layer for generating the classification results. The neuron numbers of the Softmax and the full-connection layers are 9 and 1280, respectively.

## IV. EXPERIMENTS

In this section, we first present experimental dataset and setup, and then describe experimental results and discussions.

*A. Experimental Dataset*

A publicly available dataset, which was previously adopted in DCASE-2018 Task 5, is used as experimental dataset which consists of 9 types of domestic activities. Continuous audio recording was segmented into audio segments with duration of 10 s. The audio segments are separately saved as audio files. Table I lists the numbers of audio segments for each acoustic scene contained in the training subset, validation subset, and test subset.

TABLE I. DETAILED INFORMATION OF EXPERIMENTAL DATASET.

| Domestic activities | Training no. | Validation no. | Test no. |
|---|---|---|---|
| Absence | 10001 | 2941 | 5918 |
| Cooking | 3448 | 522 | 1154 |
| Dishwashing | 916 | 110 | 398 |
| Eating | 1369 | 314 | 625 |
| Others | 1188 | 302 | 570 |
| Social activity (visit, etc.) | 3163 | 712 | 1069 |
| Vacuum cleaning | 531 | 200 | 241 |
| Watching TV | 10353 | 2331 | 5964 |
| Working (typing, etc.) | 10972 | 3120 | 4552 |
| **Total** | **41941** | **10552** | **20491** |

*B. Experimental Setup*

Classification Accuracy (CA) is used as the main performance metric, whose definition is: the number of correctly classified audio segments among the total number of audio segments [2]. Each audio segment is regarded as an independent test sample. The higher its score is, the better the performance of the method is. Another two metrics are used to measure spatiotemporal complexity of the neural networks adopted in different methods, including Parameter's Size (PS) and Multiply-Accumulate Operations (MAO). The lower the values of these two metrics are, the lower the spatiotemporal complexity of the neural network is.

We train the network using the Adam optimizer with cross-entropy loss. The learning rate is chosen to begin with 0.001 and decayed by a factor of 0.0005 if the training losses of two consecutive epochs do not decrease. All experiments are implemented with PyTorch on Google's CoLab cloud computer. Main parameters of our method are experimentally tuned and their configurations are listed in Table II.

TABLE II. MAIN PARAMETERS SETTINGS OF OUR METHOD.

| Type | Parameter's settings |
|---|---|
| Log Mel-spectrogram | Frame length/overlapping: 40ms/20ms<br>Dimension: 28 |
| Multi-scale dilated DSCN | Batch size: 32<br>Learning rate: 0.001<br>Number of DSC blocks: 8<br>Number of layers per DSC block: 3<br>Number of convolutional layers (CL): 53<br>Channel no. of CL: [3, 32, 16, 24, 32, 64, 96, 160, 320, 1280]<br>Number of scales: 3<br>Dimension of embedding: 1280<br>Neuron number of full-connection layer: 1280<br>Size of a convolutional kernel: 3×3<br>Weight decay: 0.0005<br>Momentum: 0.9<br>Dilation rate: 2 |

*C. Results and Discussions*

The first experiment is an ablation experiment for proving the contributions of MSE module and dilated convolutional operation (DCO) to the performance of the proposed framework. Experimental results obtained by three versions of our method evaluated on the experimental dataset are presented in Table III.

TABLE III. RESULTS OBTAINED BY DIFFERENT VERSIONS OF OUR METHOD.

| Methods | CA |
|---|---|
| Proposed method with DCO and MSE | 0.831 |
| Proposed method without DCO | 0.797 |
| Proposed method without MSE | 0.812 |

Our method with DCO and MSE obtains CA of 0.831 and obtains absolute gains of 0.034 and 0.019 over our method without DCO and our method without MSE, respectively. Hence, both DCO and MSE have contributed to the performance improvements of our method, and the contribution of DCO exceeds that of MSE. MSE can represent the time-frequency properties of acoustic scenes at multiple resolutions, while SCE does it at only one resolution. Hence, the representation ability of MSE is stronger and better results are obtained.

In the second experiment, the proposed method is compared with three methods based on three state-of-the-art lightweight networks (MobileNet-v1 [35], MobileNet-v2 [36], and ShuffleNet [37]), and one method based on one non-lightweight network (DenseNet [38]). Main parameters of these four baseline methods are set in accordance with the suggestions in corresponding references and optimally adjusted on the experimental dataset. Experimental results achieved by all methods are presented in Table IV.

TABLE IV. COMPARISON OF DIFFERENT METHODS FOR DOMESTIC ACTIVITIES CLASSIFICATION.

| Methods | PS | MAO | CA |
|---|---|---|---|
| MobileNet-v1 based [35] | 4.38 M | 0.34 G | 0.800 |
| MobileNet-v2 based [36] | 3.50 M | 0.33 G | 0.790 |
| ShuffleNet based [37] | 2.27 M | 0.15 G | 0.819 |
| **Proposed method** | **2.67** M | **0.44 G** | **0.831** |
| DenseNet [38] | 14.51 M | 3.33 G | 0.840 |

The proposed method achieves CA of 0.831 and obtains absolute gains of 0.031, 0.041, and 0.012 over MobileNet-v1 based method, MobileNet-v2 based method, and ShuffleNet based method, respectively. In terms of spatial complexity, the PS of the proposed method is 2.67 M which is smaller than that of the methods based on the MobileNet-v1 and the MobileNet-v2, and is slightly higher than that of the method based on the ShuffleNet. In terms of time complexity, the MAO of the proposed method is 0.44 G which is higher than that of the methods based on other three lightweight networks. Based on the results above, it can be concluded that the proposed method exceeds the methods based on state-of-the-art lightweight networks in terms of CA, and that our method obtains satisfactory results in terms of spatiotemporal complexity compared to other methods based on three lightweight networks.

On the other hand, the proposed method is inferior to the method based on non-lightweight network (DenseNet) in terms of CA, and the margin of CA between these two methods is 0.009. However, the values of both PS and MAO of the proposed method are greatly lower than that of the DenseNet based method, as shown in Table IV. Therefore, compared to the DenseNet based method, our method has the advantage of lower spatiotemporal complexity and thus is suitable to be deployed in portable terminals with limited computational resources.

To visually illustrate the classification results obtained by the proposed method on the test subset, the confusions among different classes of acoustic scenes are depicted in Fig. 2. We compare the predicted labels with the target labels, and make a confusion matrix to analyze the classification ability of the proposed method. Most audio recordings of the same class of acoustic scenes are correctly classified as their corresponding target class with few confusions among different classes of acoustic scenes. For instance, audio recordings of the *Others* (4) are spread to other classes, especially the *Absence* (0). Meanwhile, audio recordings of the *Others* and *Working* (8) are easy to be confused. The reasons for these confusions may be that: the characteristic differences among these classes of acoustic scenes are not effectively represented by the MSE and there are overlapping areas in the feature-space distribution of these acoustic scenes.

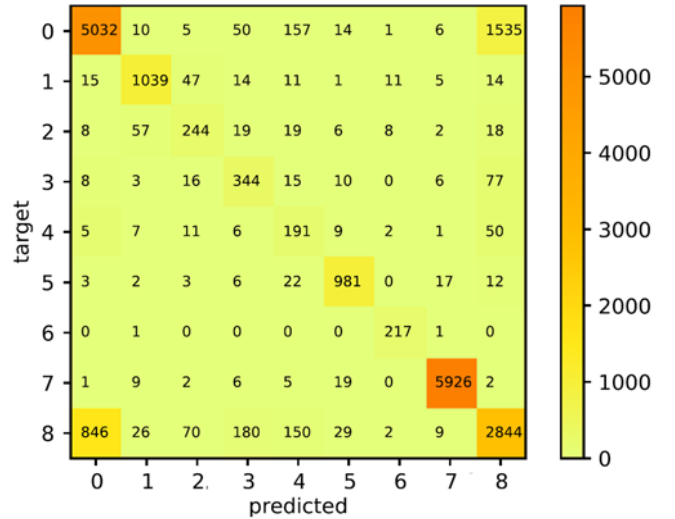

Fig. 2. Confusion matrix of the proposed method on the test subset. Digits (0 to 8) in the coordinate axes represent different classes of acoustic scenes. 0: *Absence*; 1: *Cooking*; 2: *Dishwashing*; 3: *Eating*; 4: *Others*; 5: *Social activity*; 6: *Vacuum cleaning*; 7: *Watching TV*; 8: *Working*.

## V. CONCLUSIONS

In this work, we address a problem of domestic activity classification from audio recordings by a multi-scale dilated DSCN. The proposed method outperforms the methods based on state-of-the-art lightweight networks in terms of CA. Compared with the method based on a non-lightweight network (DenseNet), the proposed method has obvious advantage in spatiotemporal complexity, but its CA is a little lower than that of the DenseNet based method. Due to the aim of this study is to implement a DAC system in the portable terminal which will be arranged in home environment for elderly care, our method has a strong advantage in this situation over the method based on non-lightweight network.

In next work, we will further improve the performance of the proposed method by exploring other structure of neural networks for feature map transformation. In addition, we will deploy the proposed method in the portable terminal for domestic activities classification and will estimate human activities from audio recordings in other situations besides home environment.